# Mechanical properties and electronic structure of the incompressible rhenium carbides and nitrides: A first-principles study

Naihua Miao,[a] Baisheng Sa,[a] Jian Zhou,[a] Zhimei Sun,[a]* and Rajeev Ahuja[b]

[a]*Department of Materials Science and Engineering, College of Materials, Xiamen University, 361005 Xiamen, People's Republic of China*

[b]*Department of Physics and Materials Science, Condensed Matter Theory Group, Uppsala University, Box 530, 75121 Uppsala, Sweden*

*Author to whom correspondence should be addressed. Tel and Fax: +86-592-2186664. Email addresses: zmsun@xmu.edu.cn and zhmsun2@yahoo.com.

## Abstract

By means of first-principles calculations, the structural stability, mechanical properties and electronic structure of the newly synthesized incompressible $Re_2C$, $Re_2N$, $Re_3N$ and an analogous compound $Re_3C$ have been investigated. Our results agree well with the available experimental and theoretical data. The proposed $Re_3C$ is shown to be energetically, mechanically and dynamically stable and also incompressible. Furthermore, it is suggested that the incompressibility of these compounds is originated from the strong covalent bonding character with the hybridization of 5*d* orbital of Re and the 2*p* orbital of C or N, and a zigzag topology of interconnected bonds, e.g., Re-Re, Re-C or Re-N bonding.

**Key words:** A. Hard materials; D. Mechanical properties; E. First-principles

## 1. Introduction

Transition-metal (TM) carbides, nitrides and borides are of great interest and importance for their useful mechanical and electrical properties. Among the TMs, rhenium receives considerable attentions due to its high bulk modulus and shear modulus. Rhenium diborides have been synthesized and broadly studied for their incompressibility and great hardness [1]. Recently, $Re_2C$ (P63/mmc) [2-4], $Re_2N$ (P63/mmc) and $Re_3N$ (P-6m2) [5] have been prepared and shown to be incompressible, which are placed for potential technological applications. Rather great bulk modulus of $B$=405±30 GPa for $Re_2C$, $B$=401±10 GPa for $Re_2N$ and $B$=395±7 GPa for $Re_3N$ have been reported [2, 3, 5], respectively, which is comparable to that of the well-known incompressible materials c-BN. The unique property of these carbides and nitrides opens an interesting topic demanding of further exploration. Among these compounds, $Re_2C$ and $Re_3N$ have been studied theoretically and experimentally [2-4, 6, 7]. However, an extended study on the physical properties of $Re_2N$ and $Re_3C$ has not been performed and that why these compounds are so incompressible is still unknown. This provides us the motivation to undertake a detail investigation on their mechanical properties and electronic structure and to understand the origin of their incompressibility.

First-principles computational technique provides us a powerful tool to explore the phase stability and physical and mechanical properties of materials. In the present work, by means of first-principles calculations, we focused on rhenium carbides and nitrides ($Re_nX$, $n$=2 or 3, X=C or N). Among them, the proposed $Re_3C$ (P-6m2) with an isostructure of $Re_3N$ has not yet been reported experimentally and its structural stability is still unknown. Hence, we began with our study on the thermodynamic stabilities of $Re_nX$, and then systematically explored their mechanical properties and electronic structure. Our results will provide a fundamental understanding on these incompressible materials and will be helpful for the further experimental and theoretical investigations on this class of materials.

## 2. Calculation Methods

The present first-principles calculations are based on the density functional theory (DFT). We use the Vienna *ab initio* simulation package (VASP) code [8] as implemented to solve the Kohn-Sham equations employing the projector augmented wave (PAW) method [9-12]. The semi-core *s* and *p* states and possibly the semi-core *d* states were treated as valence states. PAW-GGA-PBE [11, 13, 14] pseudo-potentials with electronic configurations of C $2s^22p^2$, N $2s^22p^3$, and Re $5p^66s^25d^5$ were employed. The cutoff energy for plane wave basis set was 800 eV. 13×13×4 k-points for $Re_2C$ and $Re_2N$ and 14×14×6 for $Re_3C$ and $Re_3N$ with Monkhorst-Pack (MP) scheme [15] were adopted for Brillouin zone sampling. The relaxation convergence for ions and electrons were $1\times10^{-5}$ eV and $1\times10^{-6}$ eV, respectively. The lattice parameters, force constants, elastic constants and density of states and electron localization function (ELF) [16, 17] analyzed by VESTA [18] were calculated for the equilibrium structures. The elastic constants were calculated by using the stress-strain methods. The stress tensors were calculated by VASP and then the elastic coefficients were extracted according to the Hooke's law. The phonon calculations for $Re_3C$ were performed through the supercell approach [19]. Force constants of supercells were obtained by using the VASP, and the PHONOPY code [20, 21] was performed to calculate the phonon frequencies and phonon density of states.

## 3. Results and Discussion

The calculated lattice parameters of $Re_2C$, $Re_2N$, $Re_3N$ and $Re_3C$ are given in Table 1. It is noted that the lattice parameters predicted by VASP are in good agreement with the available experimental data for $Re_2C$, $Re_2N$ and $Re_3N$ [3, 5] and the theoretical results for $Re_2C$ [6]. The calculated values of $a$ and $c$ are around 1% larger than the experimental ones which is within the GGA overestimated error. Moreover, the predicted $z$ constants also coincide well with the experiments [3, 5]. For $Re_3C$, there is no available experimental or theoretical data, hence our results can serve as a prediction for future investigations.

To investigate the thermodynamic stability of these rhenium carbides and nitrides ($Re_nX$, $n$=2 or 3, X=C or N), we have calculated the formation energy $E_{form}$ by Eq. (1) according to the reaction (2).

$$E_{form} = E_{total}^{(Re_nX)} - nE_{total}^{(Re)} - E_{total}^{(X)} \qquad (1)$$

$$n\mathrm{Re} + \mathrm{X} \rightarrow \mathrm{Re}_n\mathrm{X} \qquad (2)$$

The total energy $E_{total}$ of Re, C and N were calculated for stable crystalline rhenium (P63/mmc), diamond (Fd-3m), and alpha dinitrogen (Pa-3), respectively. The results are also given in Table 1. For all the studied compounds, including the proposed compound $Re_3C$, small formation energies are observed, suggesting that all of them are thermodynamically stable under certain conditions (e.g., high temperature and high pressure [2, 3, 5]). It is worth noting that the positive $E_{form}$ value 43.14 meV/atom for $Re_2N$ points out that the synthesis of $Re_2N$ through reaction (2) is not a self-driven process. Nevertheless, such a tiny positive value of formation energy indicates that the reaction can be easily driven by a certain temperature or pressure. Among these compounds, $Re_2C$, $Re_2N$ and $Re_3N$ have been prepared by experiments [2, 3, 5], as the negative $E_{form}$ value -74.24meV/atom for $Re_3C$ is calculated, it is expected that $Re_3C$ could be also obtained by experiments. Moreover, a phonon dispersion calculation along the high symmetry directions has been performed to further explore the dynamical stability of the new predicted compounds $Re_3C$, which is illustrated in Fig. 1. It is known that imaginary frequencies indicate the dynamical instability of crystals. As seen from the phonon dispersion curves in Fig. 1, no negative frequency have been found for $Re_3C$, suggesting it is dynamically stable.

For hexagonal crystals, there are five independent elastic stiffness constants. The shear modulus ($G$) and the bulk modulus ($B$) according to Voigt approximations [22] are defined as: $G = (2c_{11}+c_{33}-c_{12}-2c_{13}+6c_{44}+3c_{66})/15$; $B = (2c_{11}+2c_{12}+c_{33}+4c_{13})/9$. Then the Young's modulus ($E$), and Poisson's ratio ($v$) are calculated by: $E = 9BG/(3B+G)$; $v = (3B-2G)/2(3B+G)$. Based on these equations, we derived the corresponding values which are presented in Table. 2. The Born mechanical stability criteria for hexagonal crystal is given as: $c_{44}>0$, $c_{11}-c_{12}>0$, $(c_{11}+c_{12})c_{33} >2c_{13}^2$. It is obvious that all of the studied rhenium carbides and nitrides satisfy the Born criteria, hence they are all mechanically stable. As seen from Table 2, our calculated bulk modulus are in excellent agreement with the experimental values [3, 5] (where the difference are less than 0.5%), and also coincide with the theoretical values [3-7]. Moreover, the bulk modulus of these studied rhenium carbides and nitrides are greater than 390 GPa which are greater than that of c-BN (376 GPa) [23], hence all of them are considered to be incompressible materials. It is also noted in Table 2 that, for all the compounds studied here, their Poisson's ratio ($v$) are small, indicating all of them are relatively stable against shear.

To study the ductility and brittleness of $Re_nX$, we refer to the Cauchy pressure ($c_{13}$-$c_{44}$ for hexagonal crystals) and the ratio of bulk modulus to shear modulus (B/G). Generally, a positive Cauchy pressure, or a B/G value larger than 1.75, reveals damage tolerance and ductility of a crystal, while a negative Cauchy pressure, or a B/G value smaller than 1.75, demonstrates brittleness [24]. The Cauchy pressure for $Re_2C$, $Re_3C$, $Re_2N$ and $Re_3N$ are -32 GPa, -42 GPa, 98 GPa and 107 GPa, respectively.

And the calculated B/G for $Re_2C$, $Re_3C$, $Re_2N$ and $Re_3N$ are 1.55, 1.56, 1.99 and 2.03, respectively. Obviously, the results of B/G and Cauchy pressure indicate the ductile nature of $Re_nC$ and the brittle nature $Re_nN$. To quantify the elastic anisotropy of $Re_nX$, we have calculated the shear anisotropic factor $A = 4c_{44} / (c_{11}+c_{33}-2c_{13})$ for the {1 0 0} shear planes between the <0 1 1> and <0 1 0> directions, which is identical to the shear anisotropy factor for the {0 1 0} shear planes between <1 0 1> and <0 0 1> directions. For an isotropic crystal, $A$ is equal to 1. The magnitude of the deviation from 1 is a measure of the degree of elastic anisotropy possessed by the crystal [25]. The calculated shear anisotropic factor $A$ for $Re_2C$, $Re_3C$, $Re_2N$ and $Re_3N$ are 0.79, 0.83, 0.90 and 0.97, respectively. Hence, $Re_nC$ are relatively more anisotropic than $Re_nN$.

To explore the origin of incompressibility of $Re_nX$ and gain an understanding of their electronic structure and chemical bonding, we have calculated the electronic density of states (DOS) and electron localization function (ELF). Either the DOS or the ELF of $Re_nC$ and $Re_nN$ show similar characters, hence we took the DOS plots of $Re_2N$ and the ELF plots of $Re_2N$ and $Re_3N$ for further discussion, which have been presented in Fig. 2 and Fig. 3. Note that there are finite values at the Fermi levels of the DOS in Fig. 2, indicating all of them display a metallic conductivity. The typical feature of the DOS in Fig. 2 is the presence of the so called pseudo-gap, which is the borderline of the bonding and anti-bonding states, suggesting that a strong covalent bonding character in these compounds. From the partial density of states (PDOS) in Fig. 2, it is observed that the hybridization of 5$d$ orbital of Re and the 2$p$ orbital of N (or C) contribute to the Re-Re and Re-N (or Re-C) bonding. As seen from Fig. 3, along the c-axis, there are zigzag topology of interconnected bonds of -Re-N-Re-Re-N-Re- for $Re_2N$ in Fig. 3 (a) and -N-Re-Re-Re-N- for $Re_3N$ in Fig. 3 (b), respectively.

Hence, it is clear that along the c-axis of $Re_nN$, the interconnected bonding strengths are mainly attributing to the hybridization of 5*d*-2*p* (Re-N bonding) and 5*d*-5*d* (Re-Re bonding). Moreover, it can be seen in Fig. 3 (a) and (b) that the ELF value of Re-N bonding are greater than 0.75, suggesting a strong covalent bonding character, while the ELF value of Re-Re bonding are around 0.50, indicating a typical metallic Re-Re interatomic bonding. The similar character can be observed in the ELF of $Re_nC$. Therefore, it is evident that these strong directional bonding chains with a zigzag topology account for the incompressibility of $Re_nN$ ($Re_nC$), which is quite similar to that in the incompressibility of 5d transition-metal diborides [26].

## 4. Conclusions

In summary, the structural stability, mechanical properties and electronic structure of rhenium carbides and nitrides have been systematically studied by means of first-principles calculations. The calculated lattice parameters and elastic constants for them are in good agreement with the available experimental data. All the studied compounds are shown to be incompressible with bulk modulus greater than 390 GPa. The calculated density of states of $Re_nX$ indicates that all these borides display a metallic conductivity. Furthermore, our analysis on their electronic structure and electron localization function suggests that the incompressibility of these compounds mainly attributes to strong covalent bonding character with the hybridization of 5*d* orbital of Re and the 2*p* orbital of C or N, and a zigzag topology of interconnected bonds, e.g., Re-Re, Re-C or Re-N bonding. For the proposed compound $Re_3C$ which has been demonstrated to be energetically, mechanically and dynamically stable, future experimental works are recommended for further confirmation.

## Acknowledgements

This work is supported by National Natural Science Foundation of China (60976005), the Outstanding Young Scientists Foundation of Fujian Province of China (2010J06018) and the program for New Century Excellent Talents in University (NCET-08-0474). The State Key Laboratory for Physical Chemistry of Solid Surfaces at Xiamen University is greatly acknowledged for providing the computing resources.

**Table 1.** The calculated lattice parameters ($a$, $c$ and $z$) and formation energy $E_{\mathrm{form}}$ (meV/atom) for $Re_nX$. Calc., Exp. and Theo. represent the calculated values in the present work, the experimental and theoretical results, respectively.

| $Re_nX$ | $a$(Å) | | | c(Å) | | | $z$ | | $E_{\mathrm{form}}$ |
|---|---|---|---|---|---|---|---|---|---|
| | Calc. | Exp. | Theo. | Calc. | Exp. | Theo. | Calc. | Exp. | |
| $Re_2C$ | 2.866 | 2.836[a] | 2.862[c] | 9.931 | 9.86[a] | 9.903[c] | 0.109 | 0.111[a] | -92.482 |
| $Re_2N$ | 2.864 | 2.844[b] | 2.837[b] | 9.892 | 9.796[b] | 9.799[b] | 0.106 | 0.106[b] | 43.139 |
| $Re_3C$ | 2.849 | - | - | 7.181 | - | - | 0.195 | - | -74.237 |
| $Re_3N$ | 2.835 | 2.811[b] | 2.825[b] | 7.202 | 7.154[b] | 7.159[b] | 0.198 | 0.198[b] | -32.483 |

[a] Ref. [3]. [b]Ref. [5]. [c]Ref.[6].

**Table 2.** The calculated elastic stiffness constants $c_{ij}$ (in GPa) bulk modulus ($B$), shear modulus ($G$), Young's modulus ($E$) and Poisson's ratio ($v$) for polycrystalline $Re_nX$ aggregates.

| $Re_nX$ | $c_{11}$ | $c_{12}$ | $c_{13}$ | $c_{33}$ | $c_{44}$ | $B$(GPa) | | | $G$(GPa) | $E$(GPa) | $v$ |
|---|---|---|---|---|---|---|---|---|---|---|---|
| | | | | | | Calc. | Exp. | Theo. | | | |
| $Re_2C$ | 717 | 219 | 209 | 927 | 241 | 404 | 405[a] | 385[a], 389[c], 400[d] | 261 | 644 | 0.23 |
| $Re_2N$ | 625 | 204 | 282 | 755 | 184 | 394 | 401[b] | 415[b] | 198 | 508 | 0.28 |
| $Re_3C$ | 695 | 230 | 200 | 877 | 242 | 392 | - | - | 252 | 624 | 0.23 |
| $Re_3N$ | 618 | 203 | 290 | 719 | 183 | 391 | 395[b] | 413[e] | 193 | 497 | 0.29 |

[a] Ref. [3]. [b]Ref. [5]. [c]Ref.[4]. [d]Ref.[6]. [e]Ref.[7].

**Figure caption:**

**Fig. 1.** The calculated phonon dispersion curve for $Re_3C$.

**Fig. 2.** (Color online) The calculated density of states for (a) $Re_2C$, (b) $Re_3C$, (c) $Re_2N$, (d) $Re_3N$. The Fermi levels have been set to 0 eV and marked by short dash lines.

**Fig. 3.** (Color online) Structures and contour plots of ELF on the (110) plane of $Re_nN$ for (a) $Re_2N$, (b) $Re_3N$. The color scale for the ELF value is given at the bottom of the figure, where all the mappings are under the same saturation levels and the interval between two nearest contour lines is 0.13.